# Purification and recovery of $^{100}$MoO$_3$ in crystal production for AMoRE experiment

O. Gileva,[a,*] J.S. Choe,[a] H.J. Kim,[b] Y.D. Kim,[a,c] M.H. Lee,[a,b] H.K. Park,[d] and K.A. Shin[a]

[a] *Center for Underground Physics, Institute for Basic Science (IBS),
    Daejeon, 34126, Korea*

[b] *Department of Physics, Kyungpook National University,
    Daegu, 41566, Korea*

[c] *IBS School, University of Science and Technology (UST),
    Daejeon, 34113, Korea*

[d] *Department of Accelerator Science, Korea University,
    Sejong, 30019, Korea*

   *E-mail*: `gilyova_ov@mail.ru, gilevaolga@ibs.re.kr`

ABSTRACT: The AMoRE collaboration searches for the neutrinoless double-beta decay of $^{100}$Mo with ultra-radiopure molybdate crystals operated as low-temperature scintillating bolometers. For such a rare event search experiment, the techniques for investigating and reducing radioactive background contaminants in detector materials are extremely crucial. This paper discusses techniques for deep purification of enriched molybdenum for growing the crystals and the recovery of $^{100}$MoO$_3$ from the residual melt left after growing lithium molybdate crystals. The purities of enriched molybdenum trioxide powders before and after the purification and that of the recovered powder were tested with ICP-MS; results of these tests are presented.

---

[*] Corresponding author

# Contents

## 1. Introduction

The Advanced Mo-based Rare process Experiment (AMoRE) searches for neutrinoless double-beta (0νββ) decay of $^{100}$Mo with molybdate crystals operating as low-temperature scintillating bolometers [1]. Currently, several molybdenum-containing crystals, such as $^{40}$Ca$^{100}$MoO$_4$, Li$_2$$^{100}$MoO$_4$, and Na$_2$$^{100}$Mo$_2$O$_7$, are being considered as detector element candidates for the next stage of the experiment, AMoRE-II [2–6]. For rare event search experiments like AMoRE, techniques for investigating and reducing radioactive background contaminants are extremely critical [7–9]. The first step in developing highly radiopure scintillating crystal is deep purification of the raw materials used to grow them (MoO$_3$ and carbonates of Ca, Li or Na), and its validation with precise and high-sensitivity radio-assay analyses.

The required sensitivity of the AMoRE-II detector dictates strict specifications on the internal radioactive background of the crystals. The major sources of this background are decay products of natural isotopes such as Ra, Th, and U that are primary impurities in the detector materials [10]. For the initial powders used for crystal production, the Th and U concentrations, which are measured with ICP-MS, must be below 10 ppt, and the Ra activity, which is measured with HPGe, at the μBq/kg level.

For the main crystal component, MoO$_3$, the purification technique used at Center for Underground Physics (CUP) was developed first with natural MoO$_3$ and consists of a sequence of vacuum sublimation, co-precipitation, and complete precipitation of ammonium poly-molybdates (APM) from an acidic solution [11, 12]. By the end of 2021, ~100 kg of the enriched MoO$_3$ powder will be purified using this technique. As the purification methods were applied to isotopically enriched $^{100}$MoO$_3$, the purity of the produced powder and the stepwise reduction of the impurities were continuously monitored with the ICP-MS and HPGe systems [13].

In addition to the purification of the initial powder, the residual materials that remain after crystal production must be reconstituted in order to recover and repurify enriched $^{100}$MoO$_3$. While extraction and purification of $^{100}$MoO$_3$ from the calcium molybdate melt were studied in previous R&D work [14], the main attention is paid to the recovery of enriched $^{100}$MoO$_3$ from a lithium molybdate melt in this study.

## 2. Materials and equipment

Enriched molybdenum trioxide powder, 99.997% purity grade, was produced by the JSC Production Association Electrochemical Plant (ECP) in Zelenogorsk, Russia. The $^{100}$MoO$_3$ powder was enriched to 96% in the $^{100}$Mo isotope using a gas centrifuge technique. The lithium carbonate powder (99.998% purity grade) was purchased from Alfa Aesar.

Ammonium hydroxide solution (~25% Puriss), produced by Sigma Aldrich, with high purity deionized (DI) water (resistivity of 18.2 MΩ·cm) obtained from a Milli-Q water purification system, was used for dissolving, sample-preparation, and solutions. Hydrochloric acid, ACS reagent, ca. 37% solution in water, was used after a sub-boiling purification in a Savillex DST-4000 acid purification system.

The apparatus for the sublimation and annealing the $^{100}$MoO$_3$ powder was designed in-house [12]. The purities of the initial materials and final products were monitored with an inductively coupled plasma mass spectrometry (ICP-MS) system equipped with a collision cell (Agilent 7900).

The purification, recovery, and analysis were performed in class 1,000 clean rooms at CUP.

## 3. Purification of raw $^{100}$MoO$_3$

The initial materials used for the production of high-quality radiopure molybdate-crystals for the AMoRE-II detector must be extremely radiopure. HPGe measurements of Ra activity in the powder, performed at Yangyang underground laboratory in Korea, have a detection limit of about 1.0 mBq/kg that was established by a one-month test run with a 1 kg sample. ICP-MS tests of the raw $^{100}$MoO$_3$ materials provide measurements of the initial contamination levels of Th, U, K, and etc. The initial contamination levels are above our requirements and, thus, the powder has to be purified before it can be used for crystal growing.

Initially, a purification technique with minimal losses was developed and validated using natural molybdenum trioxide powder as described in detail in previous publications [9, 10]. After this development, the adopted scheme was applied to the isotopically-enriched powder.

The molybdenum purification scheme that was adopted at CUP has a sequence of steps: single sublimation under low vacuum; co-precipitation with a high purity Ca-based agent; and complete precipitation of ammonium poly-molybdates (APM) from an acidic solution. In brief, the raw $^{100}$MoO$_3$ powder is sublimed at a temperature of 720°C and a vacuum pressure below 10 mTorr, after which the desublimated crystalline $^{100}$MoO$_3$ is completely dissolved in aqueous ammonium hydroxide forming an ammonium molybdate (AM) solution. A calcium chloride solution, ~1 mol% of Mo, is mixed with the AM solution at pH 8 – 9 and white calcium molybdate precipitates appear initiating co-precipitation of impurities. This pH range provides favorable conditions for co-precipitation of Th and U, and the presence of calcium-based agents indicate that radium is also removed. It should be noted that the calcium used in the process was first purified with Ra-selective sorbents to avoid cross-contamination. After filtering out the sediments, the pH of the AM solution is decreased with strong HCl to 1.6 – 1.8 while stirring continuously until APM powder precipitates, leaving the remaining Ca and other impurities in the mother solution. The extracted APM powder is subsequently annealed in an air environment into molybdenum trioxide. Step-wise purification results are presented in Table 1. Barium, which is the same family in the periodic table as radium, is measured as an indicator of the level of radium reduction.

It is important to note that the purity of the original enriched $MoO_3$ powder differs from many other natural molybdenum trioxide products because the initial contamination levels of Th and U for this powder are at the ppt level, which makes its purification challenging. The single-sublimation step at the beginning, provides efficient removal of most of the impurities, with the notable exception of lead, which sublimes simultaneously with $^{100}MoO_3$. The Th and U concentration levels are reduced by an order of magnitude, Ba and other impurities are decreased by factors that range from 2 to 6.

**Table 1.** Impurity levels of $^{100}MoO_3$ powder before and after purification. The standard uncertainties of ICP-MS measurement are estimated to be 20% for Al, K, Fe, and Ni; 10% for Sr, Ba, and Pb and 15% for Th and U; concentration levels were calculated using a standard addition calibration.

|  | Al [ppb] | K [ppb] | Fe [ppb] | Ni [ppb] | Sr [ppt] | Ba [ppt] | Pb [ppt] | Th [ppt] | U [ppt] |
|---|---|---|---|---|---|---|---|---|---|
| **Initial $^{100}MoO_3$** | 1889 | 1238 | 332 | 95 | 3034 | 18007 | 6986 | 87 | 214 |
| **Single sublimation** | 316 | 589 | 153 | <20 | 502 | 8692 | 6546 | <7.0 | 24.7 |
| **Single sublimation + co-precipitation** | 314 | 561 | 694 | <20 | <300 | 7903 | <500 | <7.0 | <8.0 |
| **Final $^{100}MoO_3$ (powder)** | 183 | 663 | 661 | <20 | <300 | 10769 | <500 | <7.0 | <8.0 |
| **Reduction factor for the final product** | 10 | 2 | 0.5 | >5 | >10 | 2 | >14 | >12 | >27 |

Based on our previous R&D work, co-precipitation is a necessary step for lead removal. It also provides additional Th and U decontamination as confirmed by the results of analysis; Pb, Th, and U concentration levels are decreased below our detection limits. The iron contamination level observed after the co-precipitation step is not significant and its final value does not exceed ppm level. After the final steps, i.e., the complete precipitation of the APM powder and its annealing, the purity of the $^{100}MoO_3$ product meets the requirements of the AMoRE experiment: Th and U are below 10 ppt, Al, K, Fe, and Pb are below ppm level, and Ni, Sr and Ba vary around the few ppb level. HPGe analysis for the radiopure materials gives upper limits and the real values for radio-activities could be only confirmed by cryogenic measurements. Moreover, the use of "wet" chemistry techniques provides a final powder with a fine homogeneous structure without any clumps, which is also a required precursor for crystal growth.

By the end of 2021, ~100 kg of the enriched powder will be purified for the AMoRE-II. With this technique, the efficiency for each run is ~ 93% for the final product, all byproducts containing residual $^{100}Mo$ are reconstituted and $^{100}MoO_3$ powder is quantitively recovered. The overall balance of $^{100}Mo$ for the process is ~ 99%.

## 4. Extraction and recovery of $^{100}MoO_3$ from LMO residual melt

After the above-described purification process, the $^{100}MoO_3$ is mixed with pure lithium carbonate powder and a $Li_2^{100}MoO_4$ (LMO) crystal is grown by a conventional Czochralski method [15]. In this method, the usual material efficiency for the crystal ingot after a single growth is ~ 35%; about 65% of the materials remain in the melt with concentrated impurities. If the initial materials are sufficiently pure before the crystal growing, the residual melt can be used second time with an added charge of powder, and the total crystal ingot yield increases to as much as 50%. After this, the residual LMO melt must be reconstituted to extract, purify, and recover the enriched molybdenum trioxide.

The LMO residual melt is dissolved in a hot DI-water with sonication and a pH of solution that is adjusted to 0.9 using strong HCl. Enriched molybdenum in the form of ammonium poly-molybdates is extracted from the LMO solution by the addition of a hot $NH_4Cl$ solution (pH ~ 0.9). For complete precipitation of the molybdenum, 1.2 moles of $NH_4Cl$ must be added per mole of Mo. After stirring for 20 min on a heating plate, the hot mixture is filtered, the filter cake is washed several times with a 10% $NH_4Cl$ solution, then it is dried and annealed to produce $^{100}MoO_3$ powder. An acidic mother solution is stored for a few days in a plastic beaker with a small crystal of APM, as a seed, to collect all of the Mo remnants. It is important for the crystal growth procedure that the remaining chloride-ions are removed from the powder after complete annealing at ~ 670 °C. The recovery efficiency for the final $^{100}MoO_3$ powder is usually ~ 98%, the remaining ~ 2% in the mother liquid is reused in the next recovery processing run.

Table 2 shows the purity of the $^{100}MoO_3$ that is recovered using the above-described method. The residual melt was obtained after a successive growth of two LMO crystals. The residual melt after the first crystal was grown was charged with fresh materials and reused. The initial $^{100}MoO_3$ in this experiment was used without any pre-treatments.

**Table 2.** ICP-MS analysis results for the $^{100}MoO_3$ recovered from the $Li_2^{100}MO$ residual melt. The uncertainties are the same as in Table 1.

| sample | Al | K | Ba | Sr | Pb | Th | U |
|---|---|---|---|---|---|---|---|
| | [ppb] | [ppb] | [ppb] | [ppt] | [ppt] | [ppt] | [ppt] |
| $Li_2^{100}MO$ residual melt | 5614 | 1388 | 62 | 3804 | 4128 | 40 | 1120 |
| Recovered $^{100}MoO_3$ | 139 | 976 | 16 | 82 | 30783 | <10 | <10 |

ICP-MS measurements verify that this recovery procedure removes all of the listed contaminants, except for Pb. The Sr and U concentrations were reduced by two orders of magnitude, Al by a factor of 40, and Ba by a factor of 4. The Th and U concentration levels were decreased to below the detection limit of 10 ppt. As was mentioned earlier for the case of Pb, a co-precipitation step is necessary. Also, in future growth runs, preliminarily purified $^{100}MoO_3$ will be used and the Pb contamination issue will not be a problem. Considering the high recovery yield of this technique, it can be successfully used to recycle the enriched molybdenum during the crystal-growing process.

## 5. Summary

The second phase of the AMoRE project aims at performing an experiment on searching the neutrinoless double-beta decay of the $^{100}Mo$ isotope using ~ 400 radiopure molybdenum-based crystals at cryogenic temperatures. This requires the development of a purification method for raw molybdenum powder and a method for Mo recovery from the residual melt of the crystal growing.

Purification techniques for $^{100}$MoO$_3$ with sequential sublimation under vacuum and "wet" chemistry (co-precipitation and complete precipitation of APM) provide a powder end-product with the required purity and efficiency.

The contaminated residual material after the growth of a Li$_2$$^{100}$MO crystal is reconstituted and the enriched $^{100}$Mo is nearly 100% recovered and purified. This recovery process enables the recycling of the unused $^{100}$MoO$_3$ powder into subsequent crystal growth processes.

## Acknowledgments

This research was funded by the Institute for Basic Science (Korea) under project code IBS-R016-D1.

## References

[1] V. Alenkovet et al., *Technical Design Report for the AMoRE 0νββ Decay Search Experiment*, arXiv:1512.05957v1 `[physics.ins-det]`.

[2] A.N. Annenkov et al, *Development of CaMoO$_4$ crystal scintillators for a double beta decay experiment with $^{100}$Mo*, Nucl. Instrum. Methods Phys. Res. A **584** (2008) 334.

[3] H.J. Kim et al., *Search for New Molybdenum-based Crystal Scintillators for the Neutrinoless Double Beta Decay Search Experiment*, Crystal Research and Technology **54** (2019) 1900079.

[4] I. R. Pandey et al., Growth and characterization of Na$_2$Mo$_2$O$_7$ crystal scintillators for rare event searches, *J. Cryst. Growth* **480** (2017) 62 .

[5] D. A. Spassky et al., *Molybdate Cryogenic Scintillators for Rare Events Search Experiments*, in Korzhik M., Gektin A. (eds) Engineering of Scintillation Materials and Radiation Technologies, ISMART 2016, January, 26-30, 2016 Minsk, Belarus. *Springer Proceedings in Physics, Springer, Cham.* 200 (2017) 242.

[6] A. Alenkov et al., *Enriched $^{40}$Ca$^{100}$MoO$_4$ Single Crystalline Material for Search of Neutrinoless Double Beta Decay,* in: Korzhik M., Gektin A. (eds) Engineering of Scintillation Materials and Radiation Technologies. ISMART 2018*,* October, 9-12, 2018 Minsk, Belarus, *Springer Proceedings in Physics, Springer, Cham.* 227 (2019) 113.

[7] A. Guliani et al., *Neutrinoless Double-Beta Decay*, Adv. High Energy Phys. **2012** (2012) 857016.

[8] R. Barnabei et al., *Crystal scintillators for low background measurements, AIP Conference Proceedings* **1549** (2013) 189.

[9] S. Dell'Oro et al., *Neutrinoless Double Beta Decay: 2015 Review*, Adv. High Energy Phys. **2016** (2016) 2162659.

[10] R. Arnold et al., *Chemical purification of molybdenum samples for the NEMO 3 experiment, Nucl. Instr. Meth. A* **474** (2001) 1.

[11] O. Gileva et al., *Investigation of the molybdenum oxide purification for the AMoRE experiment, J. Radioanal. Nucl. Chem.* **314** (2017) 3.

[12] S. Karki et al., *Reduction of radioactive elements in molybdenum trioxide powder by sublimation method and its technical performance,* 2019 JINST **14** T11002 .

[13] M.H. Lee, *Radioassay and Purification for Experiments at Y2L and Yemilab in Korea, J. Phys.: Conf. Ser.* **1468** (2020) 012249.

[14] P. Aryal et al., *Rapid simultaneous recovery and purification of calcium and molybdenum from calcium molybdate-based crystal waste, J. Mater. Cycles Waste Manag.* **21** (2019) 1384.

[15] S. J. Ra et al., *Status of ultra-pure scintillating crystal growth for rare process experiments by CUP, J. Phys.: Conf. Ser.* **1486** (2020) 012144.